\documentclass[11pt]{article}
\usepackage{Blois,epsfig}

\def\Journal#1#2#3#4{{#1} {\bf #2}, #3 (#4)}

\def\NPB{{\em Nucl. Phys.} B}
\def\PLB{{\em Phys. Lett.}  B}

\def\PRD{{\em Phys. Rev.} D}

\def\APB{{\em Acta  Phys. Polon.} B}
\def\RMP{\em Rev. Mod. Phys. }
\def\IJMPA{{\em Int. J. Mod. Phys.} A}
\def\NPBP{{\em Nucl. Phys.} B (Proc.Suppl.)}
\def\AP{\em Ann. Phys. }

\def\be{\begin{equation}}
\def\ee{\end{equation}}
\def\bea{\begin{eqnarray}}
\def\eea{\end{eqnarray}}

\newcommand{\eq}{\begin{equation}}
\newcommand{\eqx}{\end{equation}}
\newcommand{\eqn}{\begin{eqnarray}}
\newcommand{\eqnx}{\end{eqnarray}}

\begin{document}
\vspace*{2cm}
\begin{center}
\Large{\textbf{XIth International Conference on\\ Elastic and Diffractive Scattering\\ Ch\^{a}teau de Blois, France, May 15 - 20, 2005}}
\end{center}

\vspace*{2cm}
\title{VACUA OF SUPERSYMMETRIC YANG-MILLS QUANTUM MECHANICS}

\author{J. WOSIEK }

\address{M. Smoluchowski Institute of Physics, Jagellonian University, Reymonta 4 ,\\
30-059 Krakow, Poland}

\maketitle \abstracts{
Reducing Supersymmetric Yang-Mills Field
Theory to a single point in the three dimensional space results in the Supersymmetric Yang-Mills
Quantum Mechanics (SYMQM) which basically is
the effective quantum mechanics of zero momentum
modes of the original theory. Such a system is still quite non-trivial and usually inherits
many properties of the original field theory.
In this talk some beautiful features of the three-dimensional model will be reviewed and illustrated with
the aid of the recent, quantitative solution. In particular the structure of the supersymmetric
vacua and condensates will be discussed.}


Fully reduced in space field theoretical models were first
considered in 80's~\cite{CH} as simple systems with
supersymmetry~\cite{WI}. Independently, zero-volume field theories
(especially pure Yang-Mills) were employed as the starting point
of a small volume expansion, which provided an important insight
into the early lattice calculations~\cite{L,LM,VABA}.
In late 90's the models enjoyed a new interest after the
hypothesis of the equivalence, between the $D=10, SU(\infty)$
SYMQM and an M-theory of D0 branes~\cite{BFSS,BS,WAT,HS}.

The reduced, from three dimensions, quantum-mechanical Yang-Mills model is described by
nine bosonic coordinates and
momenta $x^i_a(t)$, $p^i_a(t)$, $i=1,2,3$. Fermionic degrees of freedom
are parametrizied by six creation and annihilation operators $f_a^m, f_a^{m\dagger} $ which
may be arranged into a Majorana spinor
$\psi_a^\alpha(t)$, $\alpha=1,...,4$, $m=1,2$. One can also work with Weyl spinors.
Both fermions and bosons are in the adjoint representation of the gauge group (here SU(2)), $a=1,2,3$.
The Hamiltonian
reads \cite{HS}
\eq H =  {1\over 2} p_a^ip_a^i + {g^2\over
4}\epsilon_{abc} \epsilon_{ade}x_b^i x_c^j x_d^i x_e^j + {i g
\over 2} \epsilon_{abc}\psi_a^{\dagger}\Gamma^k\psi_b x_c^k,
\label{eq:Hamiltonian} \eqx
 where $\Gamma^k$ are the standard
Dirac $\alpha^k$ matrices.

Reduced to a single point system
is still symmetric under the three dimensional rotations.
It has also the residual gauge symmetry of the original theory,
which becomes a global SU(2) rotation in a colour space.
Finally the model is invariant under the
supersymmetry transformations with
Majorana generators
\begin{equation}
Q_{\alpha}=(\Gamma^k\psi_a)_{\alpha}p^k_a + i g
\epsilon_{abc}(\Sigma^{jk}\psi_a)_{\alpha}x^j_b x^k_c. \label{QD4}
\end{equation}
The Hamiltonian (\ref{eq:Hamiltonian}) reveals two more symmetries: 1) the gauge invariant fermion number $F=f_a^{m\dagger}f_a^m$, is
conserved, and 2) the system is invariant under a particle-hole transformation.
As a consequence of 1) the Hilbert space splits into 7 sectors labeled by $F=0,1,\dots 6$, and due to 2)
the spectra in
$F$-th and $6-F$-th sectors are identical.

\noindent {\em The solution.}
An efficient method to compute the
spectrum and eigenstates of polynomial hamiltonians with a
"reasonably" large number of bosonic and fermionic variables
was developed in Refs~\cite{JW1,CW1,CW2,JW3}.
One works in the eigenbasis of the number
operators associated with all individual degrees of freedom.
Beginning with the empty (fermionic and bosonic) state, $|0>=|0_F,
0_B>$, one constructs the basis of the physical Hilbert space by acting on
$|0>$ with gauge invariant polynomials of all creation operators.
The basis is explicitly cut by restricting the total number of all
bosonic quanta~\footnote{The number of fermionic quanta at a single point in space is finite
and equals to the number of different values of the internal indices.}. Then we calculate analytically matrix
representation of the Hamiltonian and by numerical diagonalization obtain the
spectrum and the eigenstates. Satisfactory convergence with the cutoff
was observed for many rather non-trivial systems. The method works
equally well for bosons and fermions being completely insensitive
to the notorious sign problem of fermionic simulations.


\noindent{\em The spectrum - supermultiplets.}
An unusual feature of these models is the coexistence of the discrete
and continuous spectrum. For purely bosonic systems
the flat directions of the bosonic potential (\ref{eq:Hamiltonian}) are blocked by the
energy barier induced by quantum fluctuations in the transverse directions \cite{Sim}. The spectrum is discrete
with localized eigenstates sometimes referred to as zero-volume glueballs \cite{VB2}. On the other hand,
for suprsymmetric
systems fermionic and bosonic fluctuations cancel, and the continuous spectrum
appears \cite{WLN}. However the discrete spectrum also exists, and consequently
both localized and non-localized states coexist in the same energy range.
All these features were observed \cite{JW1} and confirmed with high accuracy
in Ref \cite{CW2}. Moreover, it was found that the, low energy, continuous spectrum appears only in certain channels.

\begin{figure}[ht]
\centering
 \psfig{figure=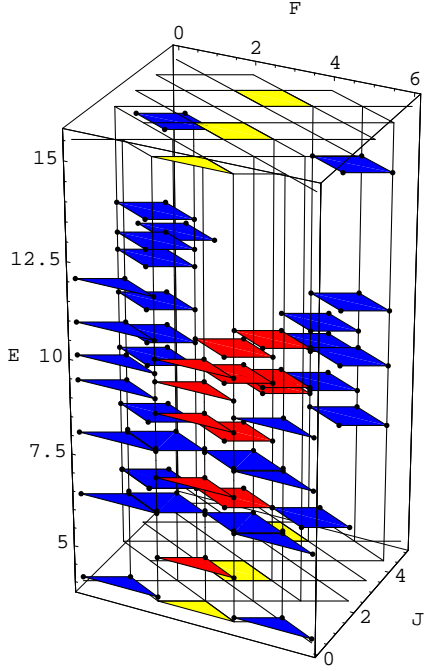,height=9cm} \caption{The spectrum, and its supersymmetry structure, of the
three dimensional supersymmetric Yang-Mills quantum mechanics.
\label{fig:3D}}
\end{figure}

Figure \ref{fig:3D}
shows a sample of eigenenergies (black dots) with various values of the fermionic number $F$ and the
angular momentum $j$. Allowed pairs $(F,j)$ define channels which are located on the vertices of
the auxilary mesh displayed on the top and the bottom of the figure.

Levels from different channels group into diamonds with the same energy.
These diamonds are nothing but  dynamical supermultiplets. They were identified on the basis
of energy degeneracies and
from the overlaps of the supersymmetric images of corresponding eigenstates.
States within supermultiplets are degenerate
to four digits, for lower levels, and up to two digits for the highest displayed
level~\footnote{For the presently available cutoffs (30-40).}.
Due to the particle-hole invariance, the diagram is symmetric with respect to the $F=3$ plane.
This is clearly seen for diamonds with low $E$ and $j$. For higher supermultiplets some particle-hole partners
have been omitted in order not to obscure this already busy plot. Only localized eigenstates shown.
Non-localized states (i.e. the ones from the continuous spectrum)\cite{TW} are found exclusively in the central,
i.e. $F=3$, supermultiplets
and only for even $j$ (marked with the yellow color at the top and the bottom of the figure).
However in these channels the localized states exist as well, and again form supermultiplets.
They are marked with the red color in the figure.

\noindent {\em Supersymmetric vacua and consensates.}
The spectrum of the scattering states extends to zero (not shown in the Figure), consequently
the SUSY vacuum lies in the continuum with $j=0$. There are two such states, with $F=2$ and $4$, hence there
are two SUSY vacua in this model. This is in correspondence with the general result for the unreduced
SU(N) field theory, which says that the number of vacua equals to $N$ \cite{NSVZ,RV}.

This analogy goes even further. In chiral/Weyl representation of Dirac matrices
our Majorana spinor reads \cite{CW2}
\begin{equation}
\psi_{Wa}^{T}=(f^{2\dagger}_a,-f^{1\dagger}_a,f^1_a,f^2_a). \label{eqweyl}
\end{equation}
It is then a simple matter to calculate the scalar and pseudoscalar invariants
\eq
S\equiv\bar{\psi} \psi= \lambda\lambda + (\lambda\lambda)^{\dagger},\;\;\;\;
P\equiv\bar{\psi} \gamma_5 \psi= -(\lambda\lambda - (\lambda\lambda)^{\dagger}),
\eqx
where
\eq
\lambda\lambda=2 f_a^{1}f_a^2,
\eqx
is nothing but the gluino condensate in Weyl notation \cite{Sh}. It annihilates two fermionic quanta, hence
the $S$ and $P$ map the two nonperturbative vacua  $|n>=|2_F,v>,|4_F,v>$ into each other.
\eq
<m|S|n>=\left( \begin{array}{cc}
                      0 & V \\
                      V^* & 0
                \end{array} \right),\;\;\
<m|P|n>=\left( \begin{array}{cc}
                      0 & -V \\
                      V^* & 0
                \end{array} \right),\;\;\;\; V\equiv <2_F,v|\lambda\lambda|4_F,v>.  \label{con1}
\eqx
Therefore one can form the two vacua, with well defined  particle-hole parity,
$|\pm>=(|2_F,v>\pm|4_F,v>)/\sqrt{2}$ with the opposite values of the average  gluino condensate.
\eq
<+|\lambda\lambda|+>=-<-|\lambda\lambda|->=\frac{1}{2} V.  \label{con2}
\eqx
The structure of Eq. (\ref{con2}) again agrees with the exact result known 
in the full (i.e. space extended) SU(2)  field theory  \cite{NSVZ,RV}.

Two comments are in order. Since the spectrum in the $(2_F,0_j)$ and $(4_F,0_j)$ channels is continuous one
has to check that there is no mixing with higher momentum states and that the matrix element V remains
non-zero in the infrared limit. This requires the scaling analysis similar to \cite{CW2}
when the dispersion relation for the scattering states was established. However the structure
of Eqs.(\ref{con1},\ref{con2} )depends only on the two-fermion nature of the $\bar{\psi} \psi$ and
$\bar{\psi} \gamma_5 \psi$ operators,
and on the particle-hole symmetry, hence the conclusion (\ref{con2}) should be true in general. Second,
non-vanishing $\bar{\psi} \psi$ condensate in full (i.e. space extended) field theory signals spontaneous
breaking of chiral symmetry which can occur only with the infinite number of degrees of freedom.
What is therefore the meaning of Eq.(\ref{con2}) in the finite system considered here?
\footnote{I thank P. van Baal for reminding a potential problem of such a formulation.} One should keep in mind
however, that in the zero-volume limit the new conservation law has emerged, namely the conservation
of the number of fermionic quanta $F$. This superselection rule splits the Hilbert space into sectors which do not communicate
during the time evolution. Consequently its role is analogous to the infinite number of degrees of freedom which provide
an impenetrable barier in the case of space extended systems.

\section*{Acknowledgments}

I would like to thank Gabriele Veneziano for many discussions.

This work is supported by the Polish Committee for Scientific Research under
grant no. PB 1P03B 02427 (2004-2007 ).

\end{document}